\documentclass[preprint,showpacs,amsmath,amssymb]{revtex4}
\usepackage{graphicx}
\usepackage{dcolumn}
\usepackage{bm}

\begin{document}
\title{Elementary electronic excitation from  a two-dimensional\\
hole gas in the presence of spin-orbit interaction}

\author{W. Xu}
\email{wen105@rsphysse.anu.edu.au} \address{Department of
Theoretical Physics\\ Research School of Physical Sciences and
Engineering\\ Australian National University\\ Canberra, ACT 0200,
Australia}

\author{L.B. Lin}
\address{Department of Physics, Sichuan University\\
Chengdu - 610064, Sichuan, China}

\begin{abstract}
We present a theoretical study of the elementary electronic
excitation associated with plasmon modes in a two-dimensional hole
gas (2DHG) in the presence of spin-orbit (SO) interaction induced
by the Rashba effect. The calculation is carried out using a
standard random-phase-approximation approach. It is found that in
such a spintronic system, plasmon excitation can be achieved via
intra- and inter-SO electronic transitions around the Fermi level.
As a result, the intra- and inter-SO plasmon modes can be
observed. More importantly, the plasmon modes induced by inter-SO
transition are optic-like and these modes can be directly applied
to identify the Rashba spin splitting in 2DHG systems through
optical measurements. The interesting features of the plasmon
excitation in a spin split 2DHG are analyzed and discussed in
details. Moreover, the results obtained for a 2DHG are compared
with those obtained for a spin-splitting 2DEG reported very
recently.

\end{abstract}


\date{\today} \maketitle \clearpage

\section{Introduction}

Progress made in realizing spin polarized electronic systems has
led to recent proposals dealing with novel electronic devices,
such as spin-transistors \cite{datt}, spin-waveguides \cite{wang},
spin-filters \cite{koga}, quantum computers \cite{ohno}, etc. In
recent years, spin-electronics (or spintronics) has become an
important and fast-growing research field in condensed matter
physics and semiconductor electronics. At present, spintronic
systems and devices have been realized on the basis of diluted
magnetic semiconductors and narrow-gap semiconductor
nanostructures. In the former case the spin degeneracy of the
carriers are lifted by the presence of an external magnetic field
and in the latter case the spin-orbit (SO) interaction (SOI) is
introduced due to the innate features of the material systems.
Currently, one of the most important aspects in the field of
spintronics is to investigate electronic systems with finite spin
splitting in the absence of the external magnetic field. It has
been realized that in narrow-gap semiconductor quantum well
structures, the higher-than-usual zero-magnetic-field spin
splitting (or spontaneous spin splitting) can be achieved by the
inversion asymmetry of the microscopic confining potential due to
the presence of the heterojunction \cite{scha}. This kind of
inversion asymmetry corresponds to an inhomogeneous surface
electric field and, hence, this kind of spin-splitting is
electrically equivalent to the Rashba spin-splitting or Rashba
effect \cite{rash}. The state-of-the-art material engineering and
micro- and nano-fabrication techniques have made it possible to
achieve experimentally observable Rashba effect in, e.g., InAs-
and InGaAs-based two-dimensional electron gas (2DEG) systems
\cite{dirk} and GaAs-based two-dimensional hole gas (2DHG) systems
\cite{wink}. In particular, it has been demonstrated very recently
that a spin split 2DHG with relatively strong Rashba effect can be
realized in a GaAs/AlGaAs heterojunction grown a nominally undoped
(311)A GaAs substrate with a weak p-type background doping
\cite{wink}. More interestingly, in such a system a back gate can
be applied to control the strength of the SOI in the device
\cite{wink}.

In recent years, the effect of SOI on electronic and transport
properties of 2DEGs and 2DHGs has been intensively studied both
experimentally and theoretically. At present, one of the most
powerful and most popularly used experimental methods to identify
the Rashba spin splitting in 2DEG and 2DHG systems is
magneto-transport measurements carried out at quantizing magnetic
fields and low-temperatures at which the Shubnikov-de Hass (SdH)
oscillations are observable \cite{dirk,wink,nitt,luo,tutu}. From
the periodicity and profile of the SdH oscillations, the carrier
density in different spin branches together with the Rashba
parameter can be determined experimentally. However, in GaAs-based
2DHG systems, because the holes are much heavier than electrons,
the magneto-transport measurements can only be applied to study
the Rashba spin-splitting in the low-density samples \cite{wink},
otherwise very high magnetic fields are required in order to
observe the SdH oscillations. The experimental data showed that a
stronger Rashba effect of the 2DHG can be achieved in a sample
with a larger hole density \cite{wink}. Thus, optical measurements
(e.g., optical absorption and transmission, Raman spectrum,
ultrafast pump-and-probe experiments, etc.) become one of the
realistic options in determining the spintronic properties in the
high-density 2DHGs. Furthermore, at present, most of the published
work in the field of spintronics is focused on electronic and
transport properties of 2DEGs and 2DHGs in the presence of SOI. In
order to apply the optical experiments to identify the Rashba
effect, to explore further applications of the spintronic systems
as optical devices and to achieve a deeper understanding of these
novel material systems, it is essential and necessary to examine
the effects of SOI on elementary electronic excitation from a
typical 2DHG and it becomes the prime motivation of the present
theoretical study.

It is well known that in an electron or a hole gas system, the
electronic transitions through spin- and charge-density
oscillations can result in a collective excitation associated with
plasmon oscillation modes. The spintronic materials can therefore
provide us an ideal device system in examining how electronic
many-body effects are affected by the SOI. In this paper, we
consider an interacting 2DHG where SOI is induced by the Rashba
effect. One of our aims is to obtain the modes of the elementary
electronic excitation such as plasmons and to examine the unique
features of these excitation modes. Very recently, the plasmon
modes induced by intra- and inter-SO electronic transition in
InGaAs-based 2DEG systems in the presence of the Rashba effect
have been studied theoretically \cite{wxu}. It was found that the
inter-SO transition can result in optic-like plasmon oscillations
in a spin-split 2DEG and these plasmon modes can be used to
identify the Rashba spin-splitting. On the basis that there is a
significant difference of the Rashba spin-splitting in
InGaAs-based 2DEGs and in GaAs-based 2DHGs, it is of value to
study the consequence of different types of SOI in different
spintronic systems such as spin-split 2DHGs. Furthermore, we would
like to take this opportunity to present more detailed theoretical
approaches used to study many-body effects in a 2DEG or 2DHG
system in the presence of SOI. The paper is organized as follows.
In Section II, we study spin-dependent carrier distribution along
with several single-particle aspects for a spin split 2DHG in a
GaAs-based structure. In Section III, the effects of the SOI on
dielectric function matrix and plasmon excitation are investigated
analytically using a simple and standard many-body theory. The
corresponding numerical results are presented and discussed in
Section IV and the concluding remarks are summarized in Section V.

\section{Single particle aspects}

It has been shown that for the case of a spin-spilt hole gas
system, such as a p-doped AlGaAs/GaAs heterostructure, the
effective SOI due to the Rashba effect can be obtained from, e.g.,
a ${\bf k}\cdot {\bf p}$ band-structure calculation \cite{wink}.
In GaAs-based quantum well structures with p-type doping, the
heavy holes dominate and the effects of light holes can be
neglected \cite{wink}. In this case the treatment of the
spintronic properties for heavy holes can follow closely those for
electrons \cite{scha,rash1}. For a typical 2DHG formed in the
$xy$-plane and its growth-direction taken along the $z$-axis in
semiconductor quantum wells, the single-particle Hamiltonian,
including the lowest order of SOI involving the heavy holes, is
given by \cite{wink,gerc}
\begin{equation}
H=\frac{{\bf p}^{2}}{2m^{\ast }}I_0+\beta_R
(\sigma_+\nabla_-^3+\sigma_-\nabla_+^3)+U(z),  \label{holeh}
\end{equation}
where $m^*$ is the hole effective mass, ${\bf p}=(p_x,p_y)$ with
$p_x=-i\hbar\partial\ /\partial x$ is the momentum operator, $I_0$
is the $2\times 2$ unit matrix, $\sigma_\pm=(\sigma_x\pm
i\sigma_y)/2$ with $\sigma_x$ and $\sigma_y$ being the Pauli spin
matrices, $\nabla_\pm=-(i\partial /\partial y\pm \partial
/\partial x)$, $U(z)$ is the confining potential of the 2DHG along
the growth direction, and $\beta_R$ is the Rashba parameter which
measures the strength of the SOI. This Hamiltonian is therefore a
$2\times 2$ matrix and the SOI does not affect the hole states
along the $z$-direction. The solution of the corresponding
Schr\"odinger equation are readily obtained, in the form of a row
vector, as
\begin{equation}
\Psi_{{\bf k}n\sigma}({\bf R})=2^{-1/2}[1, \sigma(k_x+ik_y)^3/k^3]
e^{i{\bf k}\cdot {\bf r}} \psi_n (z).  \label{h1}
\end{equation}
Here, $\sigma=\pm 1$ refers to different spin branches, ${\bf
R}=({\bf r},z)=(x,y,z)$, ${\bf k}=(k_x,k_y)$ is the hole
wavevector in the 2D plane, and $k=(k_x^2+k_y^2)^{1/2}$. The
energy spectrum of the 2DHG becomes
\begin{equation}
E_{n\sigma} ({\bf k})=E_\sigma (k)+\varepsilon_n =
{\frac{\hbar^2k^2 }{2m^*}}+\sigma\beta_R k^3+\varepsilon_n.
\label{h2}
\end{equation}
In Eqs. (\ref{h1}) and (\ref{h2}), the hole wavefunction
$\psi_n(z)$ and subband energy $\varepsilon_n$ along the
growth-direction are determined by a spin-independent
Schr\"odinger equation. From the hole energy spectrum given by Eq.
(\ref{h2}), one can immediately see that in the presence of SOI,
the energy dispersion of a 2DHG is not parabolic anymore and the
energy levels for the $\pm$ spin branches depend strongly on hole
wavevector (or momentum). These features are in sharp contrast to
those for a spin-degenerate 2DEG or 2DHG. Furthermore, it is known
\cite{wang} that for a spin-split 2DEG, the electron wavefunction
and energy spectrum are given respectively by $\Psi_{{\bf
k}n\sigma}({\bf R})=2^{-1/2}[1,i\sigma (k_x+ik_y)/k]e^{i{\bf
k}\cdot {\bf r}}\psi_n(z)$ and $E_{n\sigma}({\bf
k})=\hbar^2k^2/2m^*+\sigma\alpha_R k+\varepsilon_n$ with
$\alpha_R$ being the Rashba parameter for a 2DEG. Thus, in
contrast to a linear-in-$k$ dependence of the energy separation
between two spin branches in a spin split 2DEG, the Rashba effect
on a 2DHG results in a cubic-in-$k$ term of the hole energy
spectrum. This is the essential difference between a 2DEG and a
2DHG in the presence of SOI, which leads to different
density-of-states and, consequently, to different spintronic
properties in a 2DEG and a 2DHG.

\begin{figure}[tbp] \vskip -3.5truecm
\includegraphics*[width=110mm,height=140mm]{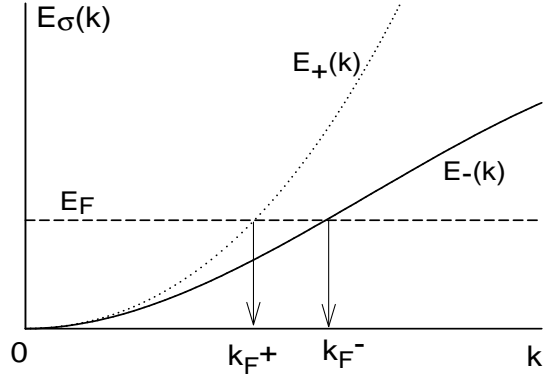}
\vspace{-4.5cm} \caption{Dispersion relation $E_\sigma (k)$  vs
$k$ for a 2DHG in different spin branches. $E_F$ (dashed line) is
the Fermi energy and the intersections of the curves for $E_\pm
(k)$ with the Fermi level, projected onto the $k$ axis, give the
Fermi wavevectors $k_F^-$ and $k_F^+$.}
\end{figure}

The dispersion relation $E_\sigma (k)$ vs $k$ resulting from Eq.
(\ref{h2}) is shown in Fig. 1. We see that when a 2DHG is in
equilibrium so that a single Fermi level ($E_F$) exists, the Fermi
wavevector in different spin branches are different, i.e., $k_F^+
< k_F^-$. This implies that the holes in the $\pm$ spin branches
have different density-of-states (DoS) below the Fermi level. In
Fig. 1 the intersections of the curves for $E_\pm (k)$ with the
Fermi level $E_F$, projected onto the $k$ axis, give the Fermi
wavevectors $k_F^-$ and $k_F^+$. The difference $k_F^--k_F^+$
leads to a difference in $k$-space area: $\pi (k_F^-)^2\ > \pi
(k_F^+)^2$. Accordingly, the hole densities in the $\pm$ branches
are different. Using hole energy spectrum given by Eq. (\ref{h2}),
the Green's function and the corresponding density-of-states (DoS)
for a 2DHG with SOI can be obtained easily. Due to energy
difference between $E_+({\bf k})$ and $E_-({\bf k})$, the DoS for
holes in the `$-$' branch is always larger than that in the `+'
branch and, as a result, the hole density in the `$-$' channel is
always larger than that in the `+' channel. Applying the DoS to
the condition of hole number conservation, for the case of a
narrow-width quantum well in which only the lowest hole subband is
present (i.e., $n=n'=0$) and at a low-temperature limit (i.e.,
$T\to 0$), the hole density $n_\sigma$ in the spin channel
$\sigma$ can be obtained by solving
\begin{equation} n_\sigma/ n_h-1/ 2+\sigma A_\beta
[(1-n_\sigma/n_h)^{3/2}+(n_\sigma/n_h)^{3/2}] =0
\label{hh4}\end{equation} for case of
$\beta_R<\hbar^2/(4m^*\sqrt{\pi n_h})$, where $n_h=n_++n_-$ is the
total hole density of the 2DHG system and $A_\beta=2 m^*
\beta_R\sqrt{\pi n_h}/\hbar^2 $. When $\beta_R\ge
\hbar^2/(4m^*\sqrt{\pi n_h})$, only the `$-$' spin branch is
occupied by holes and, therefore, the system is fully
spin-polarized (i.e., $n_+=0$ and $n_-=n_h$). However, it should
be noted that the condition $\beta_R\ge \hbar^2/(4m^*\sqrt{\pi
n_h})$ can only be satisfied in a device system with very high
hole density and very large Rashba parameter, which has not yet
been realized experimentally. Therefore, in this paper, we only
consider the situation where both $\pm$ spin branches are occupied
by holes, namely the situation where
$\beta_R<\hbar^2/(4m^*\sqrt{\pi n_h})$. To see the difference of
the carrier distribution in a 2DEG and in a 2DHG when both spin
branches are occupied, we note that for a 2DEG at a
low-temperature limit, the electron density in the $\pm$ spin
channel is given simply by \cite{wxu}: $n_{\pm}=(n_e/2)\mp
(k_\alpha/2\pi)\sqrt{2\pi n_e-k_\alpha^2}$, where $n_e=n_++n_-$ is
the total electron density and $k_\alpha=m^* \alpha_R/\hbar^2$.
These results suggest that in the presence of SOI, the spin
polarization (i.e., $(n_--n_+)/(n_-+n_+)$) increases with the
Rashba parameter for both 2DEG and 2DHG. However, for a spin-split
2DHG the spin polarization increases with {\it increasing} total
hole density, whereas the spin polarization in a 2DEG increases
with {\it decreasing} total electron density (see Figs. 1 and 2 in
Ref. \cite{wxu}).

\section{Dielectric function matrix and Plasmon modes}

\begin{figure}[tbp] \vskip -3.5truecm
\includegraphics*[width=110mm,height=140mm]{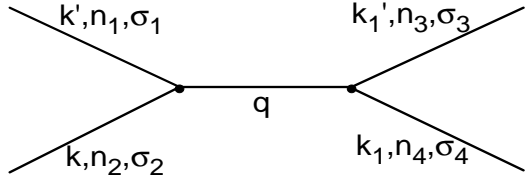}
\vspace{-6.cm} \caption{The bare hole-hole interaction in the
presence of SOI. Here ${\bf q}$ is the change of the hole
wavevector during a scattering event.}
\end{figure}

We now study many-body effects of a 2DHG in the presence of SOI.
Applying the hole wavefunction given by Eq. (\ref{h1}) to the
hole-hole (h-h) interaction Hamiltonian induced by the Coulomb
potential (see Fig. 2), the space Fourier transform of the matrix
element for bare h-h interaction is written as $$
V_{\sigma_1\sigma_2\sigma_3\sigma_4;n_1n_2n_3n_4} ({\bf k},{\bf
q})=\delta_{{\bf k}'+{\bf k},{\bf k_1}'+{\bf k_1}}V_q \int d^3{\bf
R}_1 \int d^3 {\bf R}_2 \ \Psi_{{\bf k}'n_1\sigma_1} ({\bf R}_1)
\Psi_{{\bf k}n_2\sigma_2} ({\bf R}_1)
$$
\begin{equation}
\quad\quad\quad\quad\quad\quad\quad \times e^{i{\bf q}\cdot({\bf
r_1}-{\bf r_2})} e^{-q|z_1-z_2|} \Psi_{{\bf k_1}'n_3\sigma_3}
({\bf R}_2) \Psi_{{\bf k_1}n_4\sigma_4} ({\bf R}_2)
,\label{h4}\end{equation} where ${\bf q}=(q_x,q_y)$ is the change
of the hole wavevector along the 2D-plane during a h-h scattering
event and $V_q=2\pi e^2/\kappa q$ with $\kappa$ being the
dielectric constant of the material. From now on, we consider a
narrow-width quantum well structure in which only the lowest
heavy-hole subband is present (i.e., $n'=n=0$). After defining
$\alpha=(\sigma'\sigma)$, the bare h-h interaction in the presence
of SOI becomes
\begin{equation}
V_{\alpha\beta} ({\bf k},{\bf q})=V_q F_0 (q)\Bigl[{1+\alpha
A_{\bf kq} \over 2}\delta_{\alpha,\beta}+{i\alpha B_{\bf kq}\over
2}(1-\delta_{\alpha,\beta})\Bigr], \label{h5}\end{equation} where
$F_0(q)=\int d z_1 \int d z_2\ |\psi_0(z_1)|^2 |\psi_0(z_2)|^2
e^{-q|z_1-z_2|}$ with $\psi_0(z)$ being the hole wavefunction
along the growth-direction, $A_{\bf kq}=[k^3+3k^2q{\rm
cos}\theta+3kq^2 {\rm cos}(2\theta) +q^3{\rm cos}(3\theta)]|{\bf
k}+{\bf q}|^{-3}$, $B_{\bf kq}=[3k^2q{\rm sin}\theta+3kq^2 {\rm
sin}(2\theta) +q^3{\rm sin}(3\theta)] |{\bf k}+{\bf q}|^{-3}$, and
$\theta$ is an angle between ${\bf k}$ and ${\bf q}$. It should be
noted that in contrast to a spin-degenerate 2DEG or 2DHG for which
the bare e-e or h-h interaction does not depend on ${\bf k}$
\cite{anto}, $V_{\alpha\beta} ({\bf k},{\bf q})$ for a spin-split
2DHG depends not only on ${\bf q}$ but also on {\bf k}, because
the spin splitting depends explicitly on {\bf k}. Furthermore, for
case of a 2DEG \cite{wxu}, due to different wavefunction induced
by the SOI, $A_{\bf kq}=(k+q{\rm cos}\theta)/|{\bf k}+{\bf q}|$
and $B_{\bf kq}=q\ {\rm sin}\theta/|{\bf k}+{\bf q}|$. These
results indicate that the SOI in different systems can even lead
to a different bare particle-particle interaction.

\begin{figure}[tbp]
\vspace{-1.5cm}
\includegraphics*[width=110mm,height=140mm]{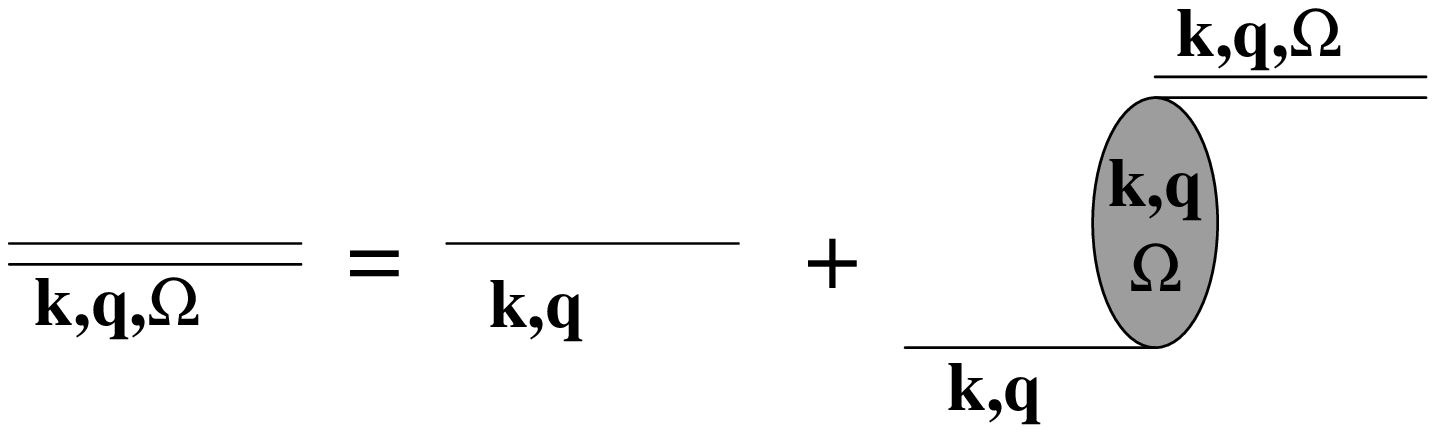}
\vspace{-8.5cm} \caption{The effective hole-hole interaction
(double solid lines) in the presence of SOI under the random-phase
approximation. Here, the single solid line is the bare h-h
interaction and the bubble refers the bare pair bubble.}
\end{figure}

From the hole energy spectrum given by Eq. (\ref{h2}), we can
derive the retarded and advanced Green functions for holes when
the effect of SOI is taken into consideration. Applying these
Green functions along with the bare h-h interaction to the
diagrammatic techniques to derive effective h-h interaction under
the random-phase approximation (RPA) (see Fig. 3), we obtain the
effective h-h interaction as
\begin{equation} V^{eff}_{\alpha\beta}(\Omega;{\bf k},{\bf q})=
V_{\alpha\beta}({\bf k},{\bf q})\epsilon_{\alpha\beta}^{-1}
(\Omega;{\bf k},{\bf q}). \label{h6}\end{equation} Here,
\begin{equation}
\epsilon_{\alpha\beta}(\Omega;{\bf k},{\bf
q})=\delta_{\alpha,\beta}\delta({\bf k})-V_{\alpha\beta}({\bf
k},{\bf q})\Pi_\beta (\Omega;{\bf k},{\bf
q})\label{h7}\end{equation} is the dynamical dielectric function
matrix element and
$$\Pi_{\sigma'\sigma}(\Omega;{\bf k},{\bf q})={f[E_{\sigma'}({\bf
k}+{\bf q})]-f[E_\sigma ({\bf k})]\over
\hbar\Omega+E_{\sigma'}({\bf k}+{\bf q})-E_\sigma ({\bf
k})+i\delta}$$ is the pair bubble or density-density correlation
function in the absence of h-h interaction, with $f(x)$ being the
Fermi-Dirac function. For a spin-split 2DHG, the effective h-h
interaction and the dielectric function matrix depend not only on
{\bf q} but also on {\bf k}, in contrast to a spin-degenerate
2DHG. After summing the dielectric function matrix over {\bf k}
and noting $\sum_{\bf k} B_{\bf kq}\Pi_{\sigma'
\sigma}(\Omega;{\bf k},{\bf q})=0$, the dielectric function matrix
for a 2DHG with Rashba spin splitting is obtained as
\begin{equation}
\epsilon=\left[\begin{array}{cccc} 1+a_1 & 0 & 0 & a_4  \\
0 & 1+a_2 & a_3 & 0 \\ 0 & a_2 & 1+a_3 & 0 \\ a_1 & 0 & 0 & 1+a_4
\end{array}\right]. \label{h8}
\end{equation}
Here, the indexes $1=(++)$, $2=(+-)$, $3=(-+)$ and $4=(--)$ are
defined regarding to different transition channels and
$a_j=-(V_qF_0(q)/2)\sum_{\bf k} (1\pm A_{\bf kq})\Pi_j
(\Omega;{\bf k},{\bf q})$ where the upper (lower) case refers to
$j=1$ or $4$ for intra-SO transition ($j=2$ or $3$ for inter-SO
transition). The determinant of the dielectric function matrix is
then given by
\begin{equation} |\epsilon|=(1+a_1+a_4)(1+a_2+a_3), \label{h9}
\end{equation}
which results from intra- and inter-SO electronic transitions.
Thus, the modes of plasmon excitation are determined by
Re$|\epsilon|\to 0$ which implies that in the presence of SOI, the
collective excitation such as plasmons can be achieved via intra-
and inter-SO transitions. It should be noted that the theoretical
approach presented here for a 2DHG can also be used for a 2DEG in
the presence of SOI and the results have been reported in Ref.
\cite{wxu}.

Because most of the conventional optical experiments measure the
long-wavelength plasmon modes. In the present work, we limit
ourselves to the case of the long-wavelength limit (i.e., $q\ll
1$). When $q\ll 1$, we have
\begin{equation} {\rm Re}a_j\simeq -{2\pi e^2 q n_\sigma \over \kappa
\Omega^2 m^*}\Bigl[1+\sigma {9\beta_R m^*\over
4\pi\hbar^2n_\sigma}\int_0^\infty dk k^2f(E_\sigma(k))\Bigr]
\label{h10}\end{equation} for intra-SO transition (i.e., for
$\sigma'=\sigma$) and
\begin{equation} {\rm Re}a_j\simeq -{9 e^2 q  \over 8\kappa}
\int_0^\infty {dk\over k} {f(E_{\sigma'}(k))-f(E_\sigma (k))\over
\hbar\Omega+(\sigma'-\sigma)\beta_R k^3} \label{h11}
\end{equation} for inter-SO transition (i.e., for
$\sigma'\neq\sigma$).

At a low-temperature limit (i.e., $T\to 0$), we have
\begin{equation} {\rm Re}(1+a_1+a_4)=1-{\omega_p^2\over \Omega^2}\Bigl( 1-
{\omega_--\omega_+\over \omega_0/2}\Bigr)\label{h12}\end{equation}
for intra-SO transition and
\begin{equation} {\rm Re}(1+a_2+a_3)=1-{\omega_p^2\over \omega_0\Omega}
{\rm ln}\Bigl({\Omega+\omega_-\over
\Omega-\omega_-}{\Omega-\omega_+\over\Omega+\omega_+}\Bigr)\label{h13}
\end{equation} for inter-SO transition. Here, $\omega_p=(2\pi e^2
n_h q/\kappa m^*)^{1/2}$ is the plasmon frequency of a
spin-degenerate 2DHG, $\omega_0=16\pi\hbar n_h/3m^*$, and
$\omega_\pm=(4\pi n_\pm)^{3/2}\beta_R/\hbar$ connected directly to
the hole density in different spin branches and to the Rashba
parameter. Thus, at the long-wavelength and low-temperature limit,
we have
\begin{equation} {\rm Re}|\epsilon|=\Bigl[1-{\omega_p^2\over \Omega^2}
\Bigl( 1-{\omega_--\omega_+\over \omega_0/2}\Bigr)\Bigr]\Bigl[1-
{\omega_p^2\over \omega_0\Omega} {\rm ln}\Bigl({\Omega+\omega_-
\over\Omega-\omega_-}{\Omega-\omega_+\over\Omega+\omega_+}\Bigr)\Bigr].
\label{h14}\end{equation} Consequently, the plasmon frequencies
induced by intra- and inter-SO excitation are given, respectively,
by
\begin{equation}\Omega_0=\omega_p
\Bigl( 1-{\omega_--\omega_+\over \omega_0/2}\Bigr)^{1/2}
\label{h15}\end{equation} and by solving \begin{equation} {\rm
ln}\Bigl({\Omega+\omega_-
\over\Omega-\omega_-}{\Omega-\omega_+\over\Omega+\omega_+}\Bigr)=
{\omega_0\Omega\over\omega_p^2}.\label{h16}\end{equation}

The theoretical results shown above indicate that in the presence
of SOI, new transition channels open up for h-h interaction and,
as a result, the collective excitation from a 2DHG can be achieved
via intra- and inter-SO transition channels. Furthermore, three
plasmon modes can be observed where one mode is caused by intra-SO
transition and two modes are induced due to inter-SO excitation.

Because the SOI results in different wavefunctions and energy
spectra in a 2DEG and in a 2DHG, although the formulas shown above
for a 2DHG look similar to those for a 2DEG \cite{wxu}, the
frequencies are defined differently. For a spin-split 2DEG,
$\omega_0=16\pi n_e\hbar/m^*$ and $\omega_\pm=4\alpha_R\sqrt{\pi
n_\pm}/\hbar$, for comparison. On the basis that $n_\pm$ for a
2DEG and for a 2DHG depends differently on the total carrier
density, $\omega_\pm$ for a 2DEG and a 2DHG has a different
dependence on the total electron and hole density.

\section{Numerical results and discussions}

Using Eq. (\ref{hh4}) we can determine the hole distribution
$n_\sigma$ in different spin branches and the plasmon frequencies
induced by different transition channels can then be calculated.
In the present study we limit ourselves to p-doped
AlGaAs/GaAs-based spintronic systems. In the numerical
calculations we take the hole effective mass $m^*=0.45 m_e$ with
$m_e$ being the electron rest-mass.

\begin{figure}[tbp]\vskip -5.5truecm
\includegraphics*[width=110mm,height=140mm]{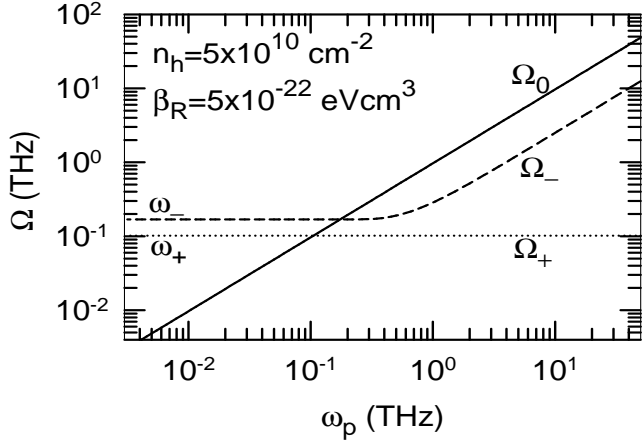}
\vspace{-1.5cm} \caption{Dispersion relation of the
long-wavelength plasmon frequency in a spin splitting 2DHG at a
fixed total hole density $n_h$ and a fixed Rashba parameter
$\beta_R$. Here, $\omega_p=(2\pi e^2 n_hq/\kappa m^*)^{1/2}\sim
q^{1/2}$, $\Omega_0$ and $\Omega_\pm$ are induced respectively by
intra- and inter-SO transitions, and $\omega_\pm=(4\pi
n_\pm)^{3/2}\beta_R/\hbar$ with $n_\pm$ being the hole density in
the $\pm$ branch.}
\end{figure}

In Fig. 4 the dispersion relation of the long-wavelength plasmon
modes induced by different transition events is shown at a fixed
total hole density and a fixed Rashba parameter. When the SOI is
present, the plasmon frequency $\Omega_0$ due to intra-SO
excitation is proportional to $\omega_p\sim q^{1/2}$ (the plasmon
frequency for a spin-degenerate 2DHG) and $\Omega_0$ is always
smaller than $\omega_p$ (see Eq. (\ref{h15})). Therefore, intra-SO
plasmons are essentially acoustic-like, similar to those in a
spin-degenerate 2DHG. In principle, the plasmon frequencies
$\Omega_\pm$ induced by inter-SO transition should depend on $q$
via $\omega_p$ (see Eq. (\ref{h16})). However, our numerical
results suggest that at a long-wavelength limit (i.e.,
$\omega_p<0.1$ THz in Fig. 4), $\Omega_\pm$ is optic-like (i.e.,
$\Omega_\pm$ depends very little on $q$) and $\Omega_\pm\to
\omega_\pm=(4\pi n_\pm)^{3/2}\beta_R/\hbar$. The most important
result shown in Fig. 4 is that at a long-wavelength limit,
inter-SO plasmons are optic-like, in sharp contrast to intra-SO
plasmons and to those observed in a spin-degenerate 2DHG. It
should be noted that $n_h\sim 10^{10}$ cm$^{-2}$ and $\beta_R\sim
10^{-22}$ eVcm$^3$ are typical sample parameters realized in
p-doped AlGaAs/GaAs quantum well structures \cite{wink}. The
results shown in Fig. 4 indicate that in these spintronic systems,
the optic-like plasmon frequencies ($\Omega_\pm$) and
$\Omega_--\Omega_+$ are within the sub-THz or high-frequency
microwave bandwidth. The similar dispersion relation of the intra-
and inter-SO plasmon modes have been observed in InGaAs-based 2DEG
systems (see Fig. 3 in Ref. \cite{wxu}). This indicates that the
SOI can result in new collective excitation modes in both 2DEG and
2DHG systems.

\begin{figure}[tbp]\vskip -3.5truecm
\includegraphics*[width=110mm,height=140mm]{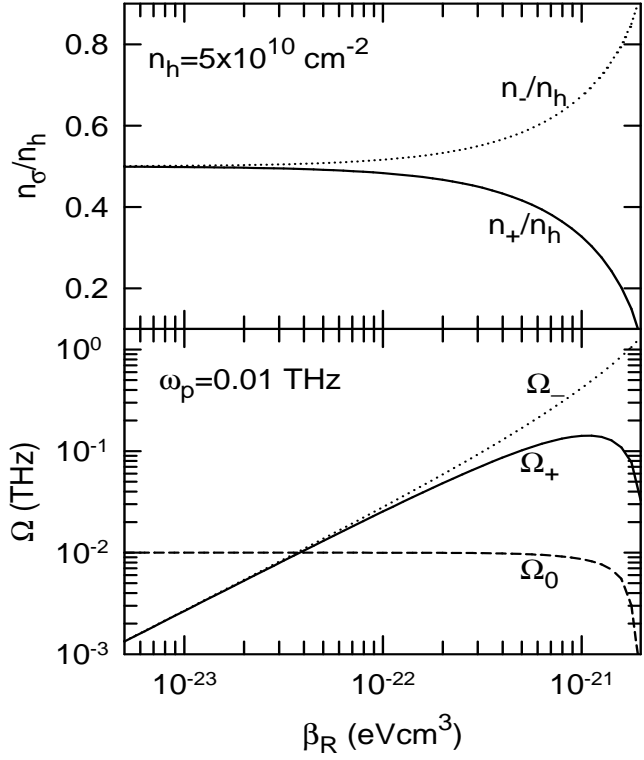}
\vspace{-0.cm} \caption{Hole density in the $\pm$ spin branches
$n_\sigma$ and plasmon frequency ($\Omega_0$ and $\Omega_\pm$
induced, respectively, by intra- and inter-SO excitation) as a
function of the Rashba parameter $\beta_R$ at a fixed total hole
density $n_h$ and $\omega_p$ as indicated.}
\end{figure}

The hole density in different spin branches and the plasmon
frequency induced by intra- and inter-SO excitation are shown in
Fig. 5 as a function of the Rashba parameter $\beta_R$ at a fixed
total hole density $n_h$ and a fixed $q$-factor via $\omega_p\sim
q^{1/2}$. With increasing the strength of the Rashba spin
splitting or $\beta_R$, more and more holes are in the `$-$' spin
branch because it has a lower energy and more DoS below the Fermi
level. As a result, with increasing $\beta_R$, because when
$\omega_p=0.01$ THz $\Omega_\pm\to\omega_\pm=(4\pi
n_\pm)^{3/2}\beta_R/\hbar$ (see Fig. 4), (i) $\Omega_-$ always
increases and $\Omega_+$ first increases then decreases; (ii)
$\Omega_-$ and $\Omega_+$ due to inter-SO excitation are more
markedly separated; (iii) the difference between $\Omega_0$
(induced by intra-SO transition) and $\omega_p$ (obtained for a
spin-degenerate 2DHG) becomes more pronounced; and (iv) at a very
large value of $\beta_R$, the plasmon excitation via intra-SO
transition can be greatly suppressed. These effects are similar to
those observed in a spin-split 2DEG (see Fig. 1 in Ref.
\cite{wxu}).

\begin{figure}[tbp]\vskip -3.5truecm
\includegraphics*[width=110mm,height=140mm]{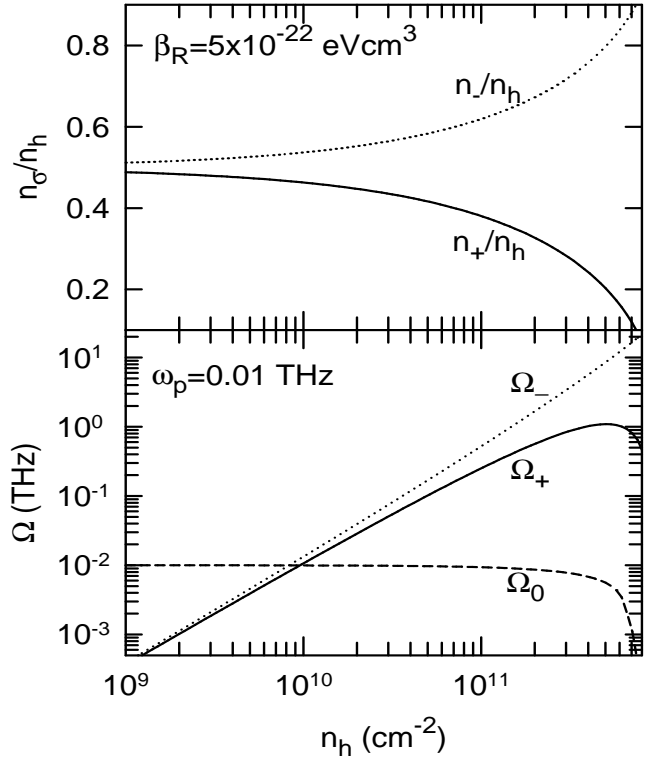}
\vspace{-0.cm} \caption{Hole distribution (upper panel) and
plasmon frequency induced by different transition channels (lower
panel) as a function of the total hole density $n_h$ at a fixed
$\beta_R$ and a fixed $\omega_p$ as indicated.}
\end{figure}

The dependence of the hole distribution and plasmon frequency due
to different transition channels on total hole density $n_h$ is
shown in Fig. 6 at a fixed $\beta_R$ and a fixed $\omega_p$. With
increasing $n_h$, due to a cubic-in-$k$ term in the energy
spectrum of the 2DHG with SOI (see Eq. (\ref{h2})), the difference
between $n_-$ and $n_+$ increases. This is in sharp contrast to
the case of a 2DEG in which $n_--n_+$ decreases with increasing
$n_e$ (see Fig. 2 in Ref. \cite{wxu}). In fact, a stronger spin
polarization achieved in a 2DHG with larger hole density has been
verified experimentally \cite{wink}. From Fig. 6, one can see that
with increasing $n_h$, (i) $\Omega_0$ induced by intra-SO
transition decreases and the excitation of this branch of plasmons
can be largely suppressed at a very high hole density; (ii) the
difference between $\Omega_\pm$ due to inter-SO excitation is
enhanced; and (iii) $\Omega_-$ always increases whereas $\Omega_+$
first increases then decreases, similar to the dependence of the
plasmon frequencies on $\beta_R$ shown in Fig. 5. Because $n_\pm$
in a 2DEG and in a 2DHG depends differently on the total carrier
density, $\Omega_\pm$ induced by inter-SO transition in a 2DHG
shows a different dependence on $n_h$ from that in a 2DEG on $n_e$
(compared Fig. 6 here to Fig. 2 in Ref. \cite{wxu}).

Since plasmon excitation from a 2DHG (2DEG) is achieved by
electronic transition around Fermi level through h-h (e-e)
interaction, changing sample parameters such as $\beta_R$
($\alpha_R$) and $n_h$ ($n_e$) implies that Fermi energy is varied
and, therefore, $\Omega_0$ and $\Omega_\pm$ depend strongly on
$\beta_R$ ($\alpha_R$) and $n_h$ ($n_e$). Moreover, a nonparabolic
subband structure of the 2DHG with SOI (see Eq. (\ref{h2}))
results in an $\Omega_0$ different from $\omega_p$ obtained from a
parabolic energy spectrum. In the presence of SOI, the electronic
transition due to h-h or e-e interaction in a 2DHG or a 2DEG has
some unique features. In a spin split 2DHG or 2DEG induced by the
Rashba effect, the spin orientation can change continuously with
the momentum orientation when a hole or an electron moves in {\bf
k}-space. Furthermore, the SOI can also shift the $\pm$ branch of
the energy spectrum continuously in {\bf k}-space, instead of a
quantized spectrum in energy space for the usual case. Thus,
conducting holes or electrons are able to change their spin
orientation simply through momentum exchange via intra- and
inter-SO transition channels due to, e.g., h-h or e-e interaction.
This process can be more easily achieved than that through energy
exchange for the usual case. Hence, although h-h or e-e
interaction is essentially an elastic scattering mechanism which
mainly alters the momentum states of holes or electrons, the
momentum variation in a spin split 2DHG or 2DEG can lead to a
significant energy exchange caused by the exchange of spin
orientation. As a result, the h-h or e-e interaction in a
semiconductor-based spintronic system can result in an efficient
momentum and energy exchange through changing the spin
orientations of the holes or electrons. Together with a
non-parabolic energy spectrum, the requirement of the momentum and
energy conversation during a h-h or e-e scattering event in a spin
split 2DHG or 2DEG differs essentially from a spin degenerate 2D
electronic system.

Normally, optical-like plasmon modes in an electronic gas system
can be measured rather conveniently by optical experiments such as
optical absorption spectroscopy \cite{apg,sja},
inelastic-resonant-light-scattering spectroscopy \cite{apg}, Raman
spectrum \cite{doa}, ultrafast pump-and-probe experiments
\cite{mvh}, etc. Acoustic-like plasmons can be detected by
applying techniques such as grating couplers to these experiments
\cite{mvh}, which in general are not so easy to measure. The
optic-like plasmon modes generated via inter-SO excitation from a
spin-split 2DHG can be detected optically. In particular, if we
can measure the long-wavelength plasmon frequencies
$\Omega_\pm\simeq \omega_\pm=(4\pi n_\pm)^{3/2}\beta_R/\hbar$ (the
magnitude of the frequencies and their separation are of the order
of 0.1 THz in a p-doped AlGaAs/GaAs heterojunction), we are able
to determine the Rashba parameter and the hole density in
different spin branches. Thus, the spintronic properties of the
device system can be obtained using optical measurements.
Especially, it is hard to use magneto-transport measurements to
identify the Rashba spin splitting in a high-density 2DHG device,
because very high magnetic fields are required to observe the SdH
oscillations. For a spin split 2DHG, larger hole density can lead
to a stronger Rashba effect (see Fig. 6) and to a more pronounced
separation between $\Omega_-$ and $\Omega_+$. Therefore, optical
experiments are very favorable in identifying the Rashba spin
splitting in high-density 2DHG samples.

On the other hand, plasmon excitation from an electronic system
can be used in realizing advanced optical devices such as light
generator and photon detector. The results obtained from this
study indicate that for p-doped AlGaAs/GaAs heterostructures, the
frequencies of the optic-like plasmons induced by inter-SO
transition can be of the order of sub-THz or high-frequency
microwaves. On the basis that in these novel material systems the
strength of the SOI can be altered by applying a back gate
\cite{wink}, the 2DHG-based spintronic systems can therefore be
applied as tunable sub-THz light generators and photon detectors.

\section{Concluding remarks}

In this paper, we have examined the effect of SOI on elementary
electronic excitation from a 2DHG in which the SOI is induced by
the Rashba effect. This work has been motivated by the recent
experimental work in which the spin split 2DHG are realized from
p-doped AlGaAs/GaAs heterostructures with relatively strong Rashba
effect. We have demonstrated that the presence of the SOI in a
2DHG can open up new channels for electronic transition via
hole-hole interaction. As a result, plasmon excitation can be
achieved via intra- and inter-SO electronic transitions, similar
to the case of a spin-split 2DEG. The interesting and important
features of these collective excitation modes have been analyzed
and been compared with those obtained for a spin-split 2DEG. The
main theoretical results obtained from this study are summarized
as follows.

In the presence of SOI, three branches of the plasmon excitation
can be generated from a 2DHG system, where one acoustic-like
branch is induced by intra-SO excitation and two optic-like
branches are due to inter-SO transition. These plasmon modes
depend strongly on sample parameters such as the total hole
density and the Rashba parameter. At a long-wavelength limit, two
optic-like plasmon modes induced by inter-SO excitation are
directly connected to the Rashba parameter and to the hole density
in different spin branches. These features are similar to those
observed in a spin-split 2DEG \cite{wxu}. Thus, the spintronic
properties in these device systems can be determined by
conventional optical measurements. Furthermore, in a p-doped
AlGaAs/GaAs heterostructure, the frequencies of the inter-SO
plasmons are of the order of sub-THz or high-frequency microwave.
These spintronic systems then can be used as optical devices such
as tunable sub-THz light generators and photon detectors.

In the presence of SOI, the hole wavefunction and energy spectrum
in a 2DHG differ significantly from those in a 2DEG. As a result,
the hole distribution in different spin branches and the
corresponding plasmon excitation modes in a 2DHG depend
differently on the total hole density from those in a 2DEG on the
total electron density. These theoretical results can be helpful
in designing optical devices based on spintronic materials.
Finally, we suggest that the important and interesting theoretical
predications merit attempts at experimental verification.

\centerline{\bf Acknowledgment} \vskip 0.5truecm

One of us (W. X.) is a Research Fellow of the Australian Research
Council. This work was also supported by the Ministry of
Education, China and National Natural Science Foundation of China.
Discussions with P. Vasilopoulos (Concordia, Canada) and M.P. Das
(ANU, Australia) are gratefully acknowledged.


\begin{references}

\bibitem{datt}  B. Datta and S. Das, Appl. Phys. Lett. {\bf 56}, 665 (1990).

\bibitem{wang}  X. F. Wang, P. Vasilopoulos, and F.M. Peeters, Phys. Rev.
{\bf B 65}, 165217 (2002).

\bibitem{koga}  T. Koga, J. Nitta, H. Takayanagi, and S. Datta, Phys. Rev.
Lett. {\bf 88}, 126601 (2002).

\bibitem{ohno} See, e.g., Y. Ohno, D.K. Young, B. Beschoten, F. Matsukura, H.
Ohno, and D.D. Awschalom, Nature {\bf 402}, 790 (1999).

\bibitem{scha} Th. Sch\"apers, G. Engels, J. Lange, Th. Klocke, M.
Hollfelder, and H. L\"uth, J. Appl. Phys. {\bf 83}, 4324 (1998).

\bibitem{rash}  E. I.  Rashba, Sov. Phys. Solid State {\bf 2}, 1109 (1960);
E. I. Rashba and V. I. Sheka, in {\it Landau Level Spectroscopy},
edited by G. Landwehr and E.I. Rashba (North-Holland, Amsterdam,
1991), Vol. 1, p.131.

\bibitem{dirk} D. Grundler, Phys. Rev. Lett. {\bf 84}, 6074 (2000), other
references therein.

\bibitem{wink}R. Winkler and U. R\"ossler, Phys. Rev. B {\bf 48}, 8918
(1993); R. Winkler, H. Noh, E. Tutuc, and M. Shayegan, Physica E
{\bf 12}, 428 (2002).

\bibitem{nitt}  J. Nitta, T. Akazaki, H. Takayanagi, and T. Enoki, Phys.
Rev. Lett. {\bf 78}, 1335 (1997).

\bibitem{luo}  J. Luo, H. Munekata, F.F. Fang, and P.J. Stiles, Phys. Rev.
{\bf B 41}, 7685 (1990).

\bibitem{tutu}  E. Tutuc, E. P. De Poortere, S. J. Papadakis, and M. Shayegan,
Phys. Rev. Lett. {\bf 86}, 2858 (2001); S.  A. Vitkalov, M.P. Sarachik, and
T. M. Klapwijk, Phys. Rev. {\bf B 64}, 73101 (2001).

\bibitem{wxu} W. Xu, Appl. Phys. Lett. {\bf 82}, 724 (2003).

\bibitem{rash1} E.I. Rashba and V. Bychkov, JETP Lett. {\bf 39}, 78
(1984).

\bibitem{gerc} Gerchikov and Subashiev, Sov. Phys. Semicond. {\bf
26}, 73 (1992).

\bibitem{anto} T. Ando, A.B. Fowler, and F. Stern, Rev. Mod. Phys.
{\bf 54}, 437 (1982).

\bibitem{apg} See, e.g., A. Pinczuk and G. Abstreiter, in {\it
Light Scattering in Solids V}, edited by M. Cardona and G.
Guntherodt (springer0Verlag, Berlin, 1989).

\bibitem{sja} See, e.g., S.J. Allen, Jr., D.C. Tsui, and
R.A.Logan, Phys. Rev. Lett. {\bf 38}, 980 (1977).

\bibitem{doa} See, e.g., D. Olego, A. Pinczuk, A.C. Gossard, and
W. Wiegmann, Phys. Rev. {\bf B 25}, 7867 (1982).

\bibitem{mvh} See, e.g., M. Vo$\beta$eb\"urger, H.G. Roskos, F.
Wolter, C. Waschke and H. Kurz, J. opt. Soc. Am. {\bf B 13}, 1045
(1996).

\end{references}
\end{document}